\documentclass[RNAAS]{aastex63}

\usepackage{amsmath}    

\newcommand{\niceurl}[1]{\href{#1}{#1}}

\begin{document}

\title{Preliminary Target Selection for the DESI Luminous Red Galaxy (LRG) Sample}



\author[0000-0001-5381-4372]{Rongpu Zhou}
\affiliation{Lawrence Berkeley National Laboratory, 1 Cyclotron Road, Berkeley, CA 94720, USA}
\affiliation{University of Pittsburgh, 100 Allen Hall, 3941 O'Hara St., Pittsburgh, PA 15260, USA}

\author{Jeffrey A. Newman}
\affiliation{University of Pittsburgh, 100 Allen Hall, 3941 O'Hara St., Pittsburgh, PA 15260, USA}

\author{Kyle S. Dawson}
\affiliation{ Department of Physics and Astronomy, University of Utah, Salt Lake City, UT 84112, USA}

\author{Daniel J.\ Eisenstein}
\affiliation{Harvard-Smithsonian Center for Astrophysics, 60 Garden St, Cambridge, MA 02138}


\author{David D. Brooks}
\affiliation{Department of Physics \& Astronomy, University College London, Gower Street, London, WC1E 6BT, UK}

\author[0000-0002-4928-4003]{Arjun Dey}
\affiliation{NSF's National Optical-Infrared Astronomy Research Laboratory, 950 N. Cherry Ave., Tucson, AZ 85719, USA}

\author[0000-0002-5665-7912]{Biprateep Dey}
\affiliation{University of Pittsburgh, 100 Allen Hall, 3941 O'Hara St., Pittsburgh, PA 15260, USA}

\author[0000-0002-2611-0895]{Yutong Duan}
\affiliation{Physics Department, Boston University, Boston, MA 02215, MA}

\author[0000-0002-8281-8388]{Sarah Eftekharzadeh}
\affiliation{Department of Physics and Astronomy, The University of Utah, 115 South 1400 East, Salt Lake City, UT 84112, USA}

\author[0000-0001-9632-0815]{Enrique Gazta\~naga}
\affiliation{Institute of Space Sciences (ICE, CSIC), 08193 Barcelona, Spain}
\affiliation{
Institut d\'~Estudis Espacials de Catalunya (IEEC), 08034 Barcelona, Spain
}

\author{Robert Kehoe}
\affiliation{Department of Physics, Southern Methodist University, 3215 Daniel Avenue, Dallas, TX 75275, USA}

\author[0000-0003-1838-8528]{Martin Landriau}
\affiliation{Lawrence Berkeley National Laboratory, 1 Cyclotron Road, Berkeley, CA 94720, USA}

\author[0000-0003-1887-1018]{Michael E. Levi}
\affiliation{Lawrence Berkeley National Laboratory, 1 Cyclotron Road, Berkeley, CA 94720, USA}

\author[0000-0002-8877-7521]{Timothy C. Licquia}
\affiliation{University of Pittsburgh, 100 Allen Hall, 3941 O'Hara St., Pittsburgh, PA 15260, USA}
\affiliation{Dow Chemical Company, 2200 W. Salzburg Road, PO Box 994, Auburn, MI 48611}

\author[0000-0002-1125-7384]{Aaron M. Meisner}
\affiliation{NSF's National Optical-Infrared Astronomy Research Laboratory, 950 N. Cherry Ave., Tucson, AZ 85719, USA}

\author[0000-0002-2733-4559]{John Moustakas}
\affiliation{Department of Physics \& Astronomy, Siena College, 515 Loudon Road, Loudonville, NY 12211, USA} 

\author{Adam D.\ Myers}
\affiliation{University of Wyoming, 1000 E. University Ave., Laramie, WY 82071, USA}

\author{Nathalie Palanque-Delabrouille}
\affiliation{IRFU, CEA, Universit\'e Paris-Saclay, F-91191 Gif-sur-Yvette, France}

\author{Claire Poppett}
\affiliation{Space Sciences Laboratory at University of California, 7 Gauss Way, Berkeley, CA 94720}

\author{Francisco Prada}
\affiliation{Instituto de Astrofisica de Andaluc\'{i}a, Glorieta de la Astronom\'{i}a, s/n, E-18008 Granada, Spain}

\author[0000-0001-5999-7923]{Anand Raichoor}
\affiliation{Institute of Physics, Laboratory of Astrophysics, Ecole Polytechnique F\'{e}d\'{e}rale de Lausanne (EPFL), Observatoire de Sauverny, 1290 Versoix, Switzerland}

\author[0000-0002-5042-5088]{David J. Schlegel}
\affiliation{Lawrence Berkeley National Laboratory, 1 Cyclotron Road, Berkeley, CA 94720, USA}

\author{Michael Schubnell}
\affiliation{Department of Physics, University of Michigan, 450 Church St., Ann Arbor, MI 48109, USA}

\author{Ryan Staten}
\affiliation{Department of Physics, Southern Methodist University, 3215 Daniel Avenue, Dallas, TX 75275, USA}

\author{Gregory Tarl\'e}
\affiliation{Department of Physics, University of Michigan, Ann Arbor, MI 48109, USA}

\author{Christophe Y\`eche}
\affiliation{IRFU, CEA, Universit\'e Paris-Saclay, F-91191 Gif-sur-Yvette, France}


\begin{abstract}

The DESI survey will observe more than 8 million candidate luminous red galaxies (LRGs) in the redshift range $0.3<z<1.0$. Here we present a preliminary version of the DESI LRG target selection developed using Legacy Surveys Data Release 8 $g$, $r$, $z$ and $W1$ photometry. 
This selection yields a sample with a uniform surface density of ${\sim}\,600$ deg$^{-2}$
and very low predicted stellar contamination and redshift failure rates. During DESI Survey Validation, updated versions of this selection will be tested and optimized. 
\end{abstract}


\section{Introduction} 

The Dark Energy Spectroscopic Instrument (DESI, \citealt{DESI2016}) will be used to obtain spectra of more than 8 million luminous red galaxies (LRGs) in the redshift range $0.3<z<1.0$, in addition to samples of stars, quasars and other galaxies. 
LRGs are massive galaxies that have typically ceased star formation and which occupy highly biased structures. Their spectra have a strong $4000\,\text{\AA}$ break, which makes their redshifts relatively easy to measure. These properties make LRGs an ideal tracer for mapping the large-scale structure of the Universe. Here, we describe a preliminary DESI LRG selection designed using Data Release 8 (DR8) of the DESI Legacy Imaging Surveys \citep{Dey2019}\footnote{\niceurl{http://legacysurvey.org/dr8/}}, and make the resulting target catalogs public\footnote{Available at \niceurl{https://data.desi.lbl.gov/public/ets/target/catalogs/} and detailed at \niceurl{https://desidatamodel.readthedocs.io}}.


\section{LRG Target Selection}

Our LRG targets are selected using Legacy Surveys (LS) DECaLS/BASS/MzLS $g, r, z$, and {\it WISE} \citep{wright_widefield_2010} $W1$ photometry \citep{Dey2019}. 
The targets are much brighter than the limits of the imaging data.  As a result the selection is insensitive to variations in imaging properties such as depth, yielding a uniform surface density across the survey footprint.

For lower-redshift LRGs, it is possible to select objects with red optical colors exploiting the 4000\,\AA~break,  as done for SDSS-I \citep{eisenstein_spectroscopic_2001a} and BOSS \citep{reid_sdssiii_2016}, but such methods fail when the break moves beyond the $r$ band.
For the eBOSS survey \citep{dawson_sdssiv_2016}, WISE photometry was used to reject stars and efficiently select massive galaxies at higher redshift (see \citealt{prakash_luminous_2015, prakash_sdss-iv_2016}).
This method takes advantage of the prominent $1.6$\,\micron\ (restframe)``bump'' \citep{John88,Sawicki02} which produces an excess of flux in the $W1$ (3.4\,\micron ) band for $z\sim1$ galaxies.  

The color and magnitude cuts used to select LRGs in the DECaLS (Southern) DR8 imaging are:
\begin{subequations}
\label{eq:cuts_south}
\begin{align}
&(z - W1) > 0.8 \times (r - z) - 0.6 , \\
&((g - W1 > 2.6) \ \mathrm{AND} \ (g - r > 1.4)) \ \mathrm{OR} \ (r - W1 > 1.8) , \\
&(r - z > (z - 16.83) \times 0.45) \ \mathrm{AND} \ (r - z > (z - 13.80) \times 0.19)  ,\\
&r - z > 0.7 ,\\
& {\rm and}\, z_{\rm fiber} < 21.5 ,
\end{align}
\end{subequations}
where $g$, $r$, $z$, and $W1$ indicate the extinction-corrected AB magnitudes in the corresponding band (using LS extinction corrections\footnote{\url{http://www.legacysurvey.org/dr8/catalogs/\#galactic-extinction-coefficients}}), and $z_{\rm fiber}$ is the magnitude corresponding to the expected $z$ flux within a DESI fiber.  
Tweaks are needed to select an equivalent sample in the Northern (BASS/MzLS) DR8 imaging due to differences in instruments and passbands; the cuts which differ are:
\begin{subequations}
\label{eq:cuts_north}
\begin{align}
&(z - W1) > 0.8 \times (r - z) - 0.65 ,\\
&((g - W1 > 2.67) \ \mathrm{AND} \ (g - r > 1.45)) \ \mathrm{OR} \ (r - W1 > 1.85) ,\\
& {\rm and}\, (r - z > (z - 16.69) \times 0.45) \ \mathrm{AND} \ (r - z > (z - 13.68) \times 0.19) .
\end{align}
\end{subequations}
%
We require all targets to be covered by at least one image in each optical band. We remove saturated objects and sources near bright stars, large galaxies, or globular clusters by requiring that LS \texttt{MASKBITS}\footnote{\url{http://www.legacysurvey.org/dr8/bitmasks/}} 1, 5, 6, 7, 11, 12, and 13 are not set.  

\begin{figure}[htb]
\centering
\includegraphics[width=16cm]{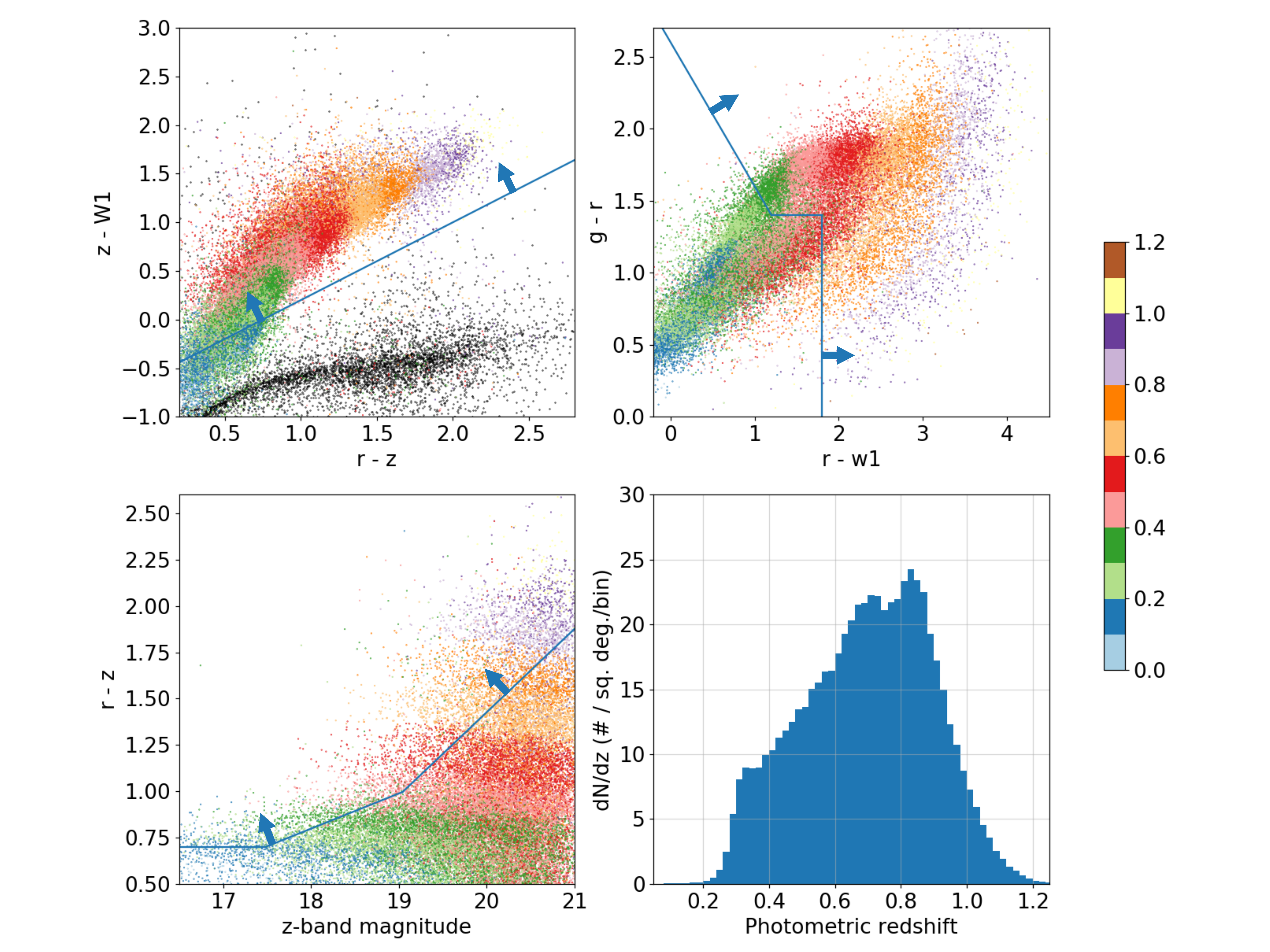}
\caption{Illustration of the selection cuts for the DESI LRG sample, as listed in Equation \ref{eq:cuts_south}.  The first three panels present optical/near-infrared color-color and color-magnitude diagrams for objects from LS DR8.
All extended sources with $z$-band fiber magnitude brighter than 21.5 are color-coded according to their PRLS photometric redshifts \citep{zhou_clustering_2020}. Blue lines depict the LRG selection boundaries given in Equation \ref{eq:cuts_south}. In the upper left panel, we also plot the color-color distribution of point sources (mostly stars) in black. The lower right panel shows the photometric redshift distribution for the selected targets, showing the number of LRGs per deg$^2$ in each $\Delta z=0.02$ redshift bin. 
}
\label{fig:lrgselection}
\end{figure}

The boundaries in color-magnitude space corresponding to Equations\,\ref{eq:cuts_south}a-e and the predicted redshift distribution for the sample are depicted in Figure \ref{fig:lrgselection}.  
The first cut 
(shown in the upper left panel of Figure \ref{fig:lrgselection}) provides an effective rejection of stars, much as in \citet{prakash_sdss-iv_2016}. 
The cuts shown in the upper-right panel eliminate low-redshift or bluer objects.  
As shown in the lower-left panel, applying a magnitude limit which is a function of color allows only the most luminous objects at a given redshift to be selected.  In addition to these cuts, we also apply a limit on $z_{\rm fiber}$ to ensure targets yield secure redshift measurements. 

We have also developed alternative approaches which apply a single, broader $r-W1$ cut independent of $g-r$, and which employ a magnitude limit in the $r-W1$ vs.\ $W1$ plane instead of in $z$. The resulting selection eliminates $g$ and reduces $z$-band dependence (making it less sensitive to photometric calibration errors)
but still yields a similar redshift distribution to that shown in Figure \ref{fig:lrgselection}.
%


\section{Conclusions} 

LRG targets selected as outlined in this Note have a total surface density of ${\sim}\,600$ deg$^{-2}$ and a roughly constant comoving density of ${\sim}\,6 \times 10^{-4}\, h^3\,\mathrm{Mpc}^{-3}$ over the redshift range $0.3<z<0.8$.
Tests with DESI commissioning data suggest that in nominal conditions $>98\%$ of these targets should yield secure redshift measurements, with $<1\%$ stellar contamination at high Galactic latitudes.  
Legacy Surveys DR9 imaging and DESI Survey Validation (SV) spectroscopy will soon be used to choose the best method for LRG targeting and optimize the selection cuts for the DESI survey. We expect the DESI LRG sample to meet or exceed all relevant Science Requirements\footnote{\url{https://cmb-s4.org/wiki/images/DESI_L123_driver.pdf}} and to surpass previous spectroscopic LRG surveys in areal coverage, redshift range, and overall number of targets.


\section{Acknowledgements}
This research is supported by the Director, Office of Science, Office of High Energy Physics of the U.S. Department of Energy under Contract No. 
DE–AC02–05CH1123, and by the National Energy Research Scientific Computing Center, a DOE Office of Science User Facility under the same 
contract, as well as by an Office of High Energy Physics grant to the University of Pittsburgh.  Additional support for DESI is provided by the U.S. National Science Foundation, Division of Astronomical Sciences under Contract No. 
AST-0950945 to the NSF’s National Optical-Infrared Astronomy Research Laboratory; the Science and Technologies Facilities Council of the United Kingdom; the Gordon 
and Betty Moore Foundation; the Heising-Simons Foundation; the French Alternative Energies and Atomic Energy Commission (CEA); 
the National Council of Science and Technology of Mexico; the Ministry of Economy of Spain, and by the DESI 
Member Institutions.  The authors are honored to be permitted to conduct astronomical research on Iolkam Du’ag (Kitt Peak), a mountain with 
particular significance to the Tohono O’odham Nation.

\bibliography{biblio}

\end{document}